\documentclass[%
 reprint,
superscriptaddress,
 amsmath,amssymb,
 longbibliography,
 prb,
]{revtex4-2}

\usepackage[ruled]{algorithm2e}
\usepackage[noend]{algpseudocode}
\usepackage{wasysym}
\usepackage{graphicx}
\usepackage{dcolumn}
\usepackage{bm}
\DeclareMathAlphabet{\mathpzc}{OT1}{pzc}{m}{it}
\usepackage{color}

\begin{document}

\preprint{APS/123-QED}

\title{NQS-Agent: Health-Aware Agentic Hyperparameter Optimization for Neural-Network Quantum States}

\author{Jia-Qi Wang} 
\affiliation{School of Physics and Key Laboratory of Quantum State Construction and Manipulation (Ministry of Education), Renmin University of China, Beijing 100872, China}

\author{Xiao-Qi Han}
\affiliation{School of Physics and Key Laboratory of Quantum State Construction and Manipulation (Ministry of Education), Renmin University of China, Beijing 100872, China}

\author{Ze-Feng Gao} 
\affiliation{School of Physics and Key Laboratory of Quantum State Construction and Manipulation (Ministry of Education), Renmin University of China, Beijing 100872, China}

\author{Rong-Qiang He} \email{rqhe@ruc.edu.cn}
\affiliation{School of Physics and Key Laboratory of Quantum State Construction and Manipulation (Ministry of Education), Renmin University of China, Beijing 100872, China}

\author{Zhong-Yi Lu}\email{zlu@ruc.edu.cn}
\affiliation{School of Physics and Key Laboratory of Quantum State Construction and Manipulation (Ministry of Education), Renmin University of China, Beijing 100872, China}
\affiliation{Hefei National Laboratory, Hefei 230088, China}

\date{\today}

\begin{abstract}
Neural-network quantum states (NQS) provide expressive variational representations for strongly correlated quantum many-body systems, but their practical accuracy depends sensitively on architecture-level hyperparameters and optimization schedules. Here we develop NQS-Agent, an implemented open-source software framework for health-aware hyperparameter optimization (HPO) in NQS calculations. Its workflow monitors energy trajectories, detects destructive optimization events, stops unstable calculations, modifies the learning-rate schedule, resumes optimization from safe checkpoints, and ranks candidates with an anomaly-aware score. We demonstrate the approach on a residual convolutional NQS for the square-lattice Heisenberg $J_1$-$J_2$ model, using architectures with parameter counts comparable to aCNN, a convolutional NQS architecture used here as a reference. The results show that NQS-Agent improves over the reported human-tuned aCNN baseline for the aCNN reference architecture and identifies a structurally distinct wide-and-shallow competitive candidate within the parameter-count-matched residual-CNN search space. These results show that the stability and recovery history of an optimization trajectory should be considered when assessing an NQS result. Health-aware HPO therefore provides a reproducible tuning protocol that goes beyond selecting a single lowest-energy calculation.
\end{abstract}


\maketitle

\section{Introduction}
\label{sec:introduction}

Strongly correlated quantum many-body systems remain a central challenge in computational physics. In frustrated magnets and other two-dimensional lattice models, the exponential growth of Hilbert space prevents direct exact treatment beyond small clusters, while the competition among local constraints, long-range correlations, and possible ordered or disordered phases makes accurate ground-state calculations especially demanding. Representative numerical approaches include quantum Monte Carlo~\cite{Sandvik1997SSEQMC}, tensor-network methods~\cite{Zheng-ChengGu2022_TensorNetwork+PEPS}, density-matrix renormalization group calculations~\cite{Stoudenmire2012_2DsystemDMRG,GongShouShu2014_DMRG,Sandvik2018_DMRG}, and variational Monte Carlo~\cite{Sorella2013_VMC}. Nevertheless, accurate solutions of intermediate-size frustrated two-dimensional systems remain difficult because the different methods face sign problems, entanglement growth, boundary effects, or large variational search spaces, respectively.

Neural-network quantum states (NQS) offer a flexible variational route to this problem by representing many-body wave functions with trainable neural architectures~\cite{GiuseppeCarleo2017_RBM,Lange2024NQSReview,Hermann2023_NQSChemistryReview,Juan2021PRX_NQSReview,Nomura2023RBMNQSReview,Medvidovic2024NQSReview}. Combined with variational Monte Carlo (VMC), NQS methods have been applied to spin models, frustrated systems, and quantum chemistry, using a growing range of architectures, including restricted Boltzmann machines~\cite{GiuseppeCarleo2017_RBM,Nomura2023RBMNQSReview}, convolutional networks~\cite{Choo2019_CNN_ComplexValued,Liang2018_CNN,fu2022latticeCNN,Jqwang2024aCNN}, recurrent networks~\cite{HibatAllah2020_RNN_NQS,Lange2024RNN_NQS}, graph-inspired networks~\cite{kochkov2021GNN,roth2023_GCNN}, and Transformers or vision-Transformer variants~\cite{23ZhangTransformerNQS,Rende2024DeepViT}. Their appeal lies in the possibility of learning compact but expressive wave-function ans\"atze directly from the variational principle.

In practice, however, the accuracy of an NQS calculation is influenced not only by the architecture class but also by architecture and optimizer hyperparameters. For convolutional NQS, the channel number, kernel size, network depth, residual-block count, learning rate, learning-rate schedule, regularization, random seed, and number of optimization steps can all influence the final energy~\cite{Jqwang2024aCNN}. Hyperparameter optimization (HPO) provides systematic ways to search these choices. Examples include random search~\cite{Bergstra2012RandomSearchHPO}, model-based search with a tree-structured Parzen estimator (TPE)~\cite{Bergstra2011AlgorithmsHPO}, and methods that evaluate many candidates briefly before assigning more optimization steps to the more promising ones, such as Hyperband~\cite{Li2018Hyperband}, the asynchronous successive halving algorithm (ASHA)~\cite{Li2020ASHA}, and BOHB, which combines Bayesian optimization with Hyperband~\cite{Falkner2018BOHB}. Software frameworks such as Optuna support the practical execution of these searches~\cite{Akiba2019Optuna}; broader overviews are available in the automated-machine-learning literature~\cite{Hutter2019AutoMLBook}. Thus, architectural choices such as depth and width and optimizer choices such as the learning-rate schedule must be tuned together under a physically meaningful comparison protocol.

On the other hand, in practice, navigating this space often requires substantial manual tuning. Expert practitioners inspect energy curves, notice abnormal jumps, decide whether an instability is transient or destructive, reduce the learning rate when the variational parameters become sensitive, and resume from a saved state when a promising calculation becomes unstable. Such interventions are feasible for a few supervised calculations, but they do not scale naturally to tens or hundreds of queued candidate experiments on shared GPU resources. This is where parallel evaluation and progressive allocation of optimization steps become useful.

This difficulty is particularly acute in NQS optimization because the energy landscape can be rugged and the training dynamics can be highly sensitive to the learning-rate scale. A small learning rate often improves stability but may slow exploration and trap the optimization in a poor local region. A large learning rate can cross shallow barriers and accelerate early descent, but it also increases the probability of upward energy jumps, non-finite values, and destructive events from which the calculation may not recover without intervention. Even when adaptive optimizers or prescribed learning-rate schedules are used, the observed energy trajectory can still become unstable and must be assessed as a time-ordered optimization curve rather than by its final value alone. This sensitivity is closely related to the broader challenge of optimizing neural quantum states with stochastic reconfiguration~\cite{Sorella2007SR}, MinSR and related natural-gradient approaches~\cite{Chen2024MinSR}, or other second-order methods~\cite{Drissi2024SecondOrderOptimization,Rende2024DeepViT}. A scalable HPO system for NQS must therefore evaluate the stability of the optimization trajectory, not merely the energy reached at a fixed evaluation point.

Here we introduce NQS-Agent, an open-source software framework for health-aware HPO that converts these manual interventions into recorded and reproducible automated operations. Recent surveys describe the broader development of agent-based scientific workflows~\cite{wei2025AIScienceAgenticScienceReview} and large-language-model multi-agent systems~\cite{Li2024surveyAgent}; MetaGPT~\cite{hong2024metagpt} and AutoGen~\cite{wu2024autogen} provide representative examples of structured multi-agent coordination. NQS-Agent combines persistent experiment state, predefined evaluation milestones, runtime health monitoring, anomaly-aware scoring, and recovery from saved states. When a destructive event is detected, the workflow stops the affected execution, identifies a safe checkpoint (a saved model and optimizer state from which optimization can continue), modifies the learning-rate schedule, and resumes the same candidate experiment instead of discarding it or restarting from the beginning.

We demonstrate the workflow on the square-lattice Heisenberg $J_1$-$J_2$ model using residual convolutional NQS based on the architecture reported in the aCNN study~\cite{Jqwang2024aCNN}. We consider architectures with parameter counts within $\pm5\%$ of the 6538-parameter aCNN architecture. From 30 architectures in this range, nine representatives, denoted A1--A9, are selected to cover different channel numbers, kernel sizes, and residual-block counts. Calculations on the $6\times6$ lattice provide an initial comparison, after which selected candidates are optimized on $10\times10$ lattices at $J_2/J_1=0.50$ and $0.55$ with runtime monitoring and checkpoint-based recovery.

The remainder of this paper is organized as follows. Section~\ref{sec:method} describes the NQS/VMC formulation, the NQS-Agent framework architecture and workflow, the diagnosis of unstable optimization, the use of saved checkpoints for recovery, and the ranking of candidates at predefined evaluation milestones. Section~\ref{sec:results} presents the aCNN benchmark, the representative parameter-count-matched architectures, representative recovered energy trajectories, and the comparison with human-tuned results. Section~\ref{sec:discussion} discusses the scope, limitations, and implications of health-aware HPO as a reproducible NQS tuning protocol.

\section{Method}
\label{sec:method}

\subsection{Neural-Network Quantum States and VMC Optimization}
\label{subsec:nqs_vmc}

An NQS represents a variational wave function by assigning a complex wave-function value $\psi_\theta(\sigma)$ to each many-body configuration $\sigma$, where $\theta$ denotes the trainable neural-network parameters. In the present work, $\sigma$ denotes a spin configuration of the square-lattice Heisenberg $J_1$-$J_2$ model. The aCNN implementation used in the numerical experiments represents the real-valued amplitude sector, but the notation $\psi_\theta(\sigma)$ is kept general because NQS ans\"atze may also encode phases or full complex wave functions.

Expectation values are evaluated by Monte Carlo sampling from the probability distribution $p_\theta(\sigma)\propto |\psi_\theta(\sigma)|^2$. For a Hamiltonian $H$, the variational energy can be written as
\begin{align}
E_\theta =
\langle H\rangle_\theta =
\mathbb{E}_{p_\theta}\left[E_{\mathrm{loc}}(\sigma)\right],
\end{align}
where the local energy is
\begin{align}
E_{\mathrm{loc}}(\sigma)=
\sum_{\sigma'} H_{\sigma\sigma'}
\frac{\psi_\theta(\sigma')}{\psi_\theta(\sigma)} .
\end{align}

The network parameters are optimized to minimize $E_\theta$. In gradient-based VMC, the energy gradient is estimated from sampled configurations as
\begin{align}
\nabla_\theta E_\theta =
2\mathrm{Re}\left[
\mathbb{E}_{p_\theta}\left[
\left(E_{\mathrm{loc}}-\mathbb{E}_{p_\theta}[E_{\mathrm{loc}}]\right)
\nabla_\theta \log \psi_\theta(\sigma)
\right]\right].
\end{align}
Stochastic reconfiguration or related natural-gradient methods can also be viewed as approximate imaginary-time evolution projected onto the variational manifold, with a parameter update controlled by a learning-rate or time-step scale~\cite{Sorella2007SR,Chen2024MinSR}.

This optimization problem is unusually sensitive because the sampled local energy, the Monte Carlo variance, the parameterization, and the learning-rate schedule interact during training. We therefore treat optimizer-level choices and structural choices as a unified HPO problem: the learning rate controls the trajectory through the variational manifold, while structural hyperparameters determine the shape and conditioning of that manifold.

\subsection{NQS-Agent Framework Architecture and Health-Aware HPO Workflow}
\label{subsec:agent_workflow}

\begin{figure*}[t!]
  \includegraphics[width=16cm]{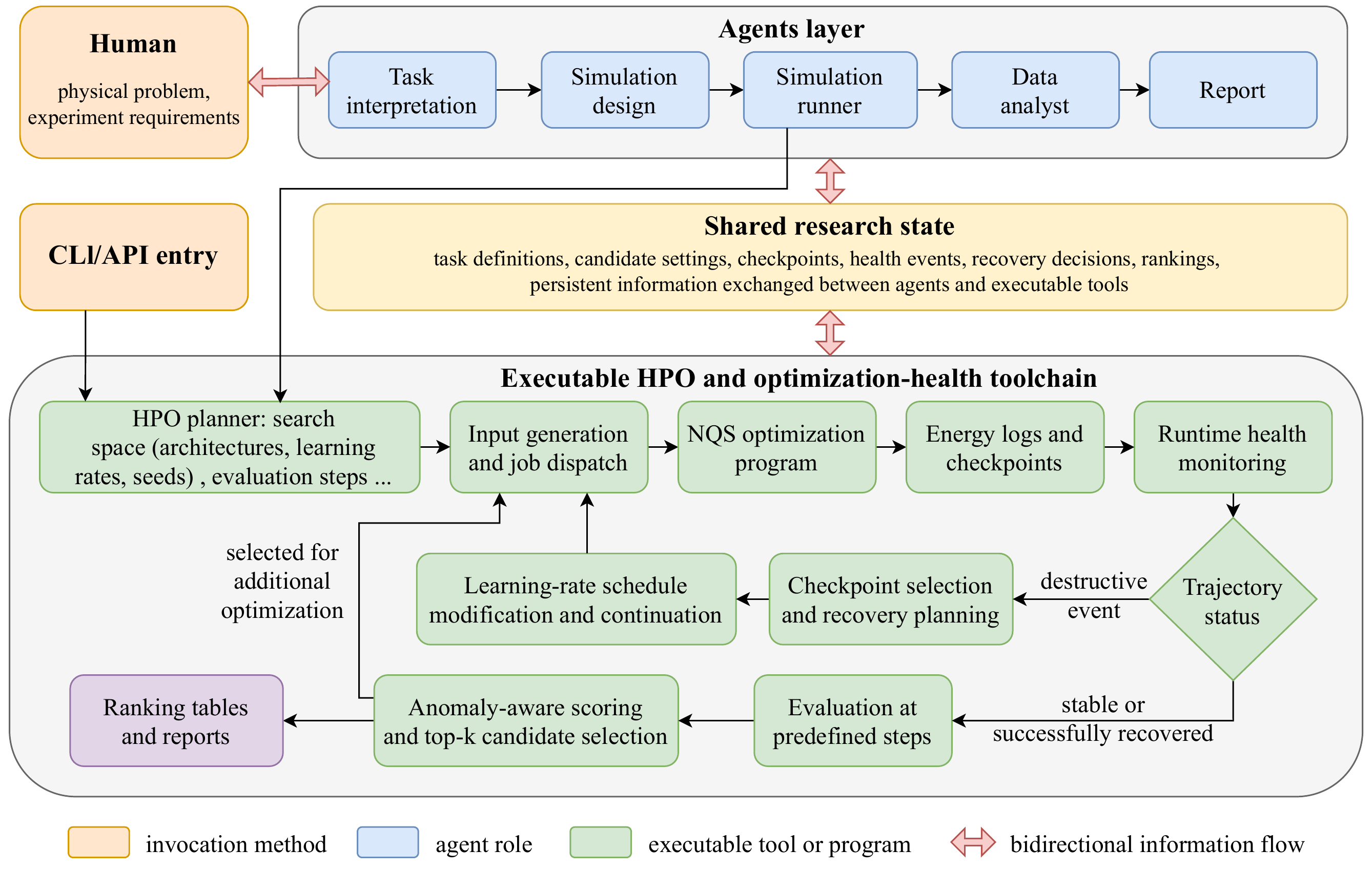}
  \caption{NQS-Agent framework architecture and task-specific health-aware HPO workflow. The agent coordination layer interprets the user task, prepares and runs experiment plans, analyzes the results, and generates reports. Task interpretation writes the task definition to the shared research state, while report generation reads the recorded state to summarize the completed study. The simulation runner invokes an executable toolchain for HPO planning, input generation, NQS optimization, runtime health monitoring, checkpoint-based recovery, anomaly-aware scoring, and top-$k$ candidate selection. The CLI/API entry provides a lower-level access path to the same deterministic tools for reproducibility checks, debugging, and user-specified batch execution, while the agent layer coordinates research-level planning, state updates, analysis, and reporting. Agents and tools exchange task definitions, candidate settings, energy logs, checkpoints, recovery decisions, and rankings through the persistent shared research state.}
  \label{fig:framework}
\end{figure*}

NQS-Agent is an implemented open-source software framework composed of an agent coordination layer, a persistent shared research state, and executable tools for HPO and optimization-health management. To be specific, NQS-Agent is implemented with LangGraph, a graph-based framework for orchestrating long-running, stateful agents~\cite{LangGraph}. The accompanying source code implements the components and interactions summarized in Fig.~\ref{fig:framework}. The agents coordinate research-level operations, whereas the executable tools perform well-defined numerical and data-management operations such as constructing candidate settings, generating simulation inputs, dispatching calculations, monitoring energy logs, selecting checkpoints, modifying learning-rate schedules, scoring candidates, and continuing selected calculations.

The agent coordination layer separates complementary roles. Task interpretation converts the user-defined objective into task definitions stored in the shared research state. Simulation design prepares the physical problems, search spaces, and experiment plans; the simulation runner invokes the corresponding executable tools. Data analysis reads the recorded energies, health events, and ranking results, and report generation uses the shared state to produce a human-readable summary. The shared state therefore connects the agent roles with the executable toolchain by storing candidate settings, resource assignments, energy logs, checkpoint information, recovery decisions, and ranking tables throughout the calculation.

The workflow distinguishes a candidate experiment from an execution segment. A candidate experiment is defined by the physical problem, network architecture, optimizer setting, and random seed. An execution segment is one uninterrupted calculation, either the initial calculation or a continuation resumed from a checkpoint. Several execution segments can therefore belong to the same candidate experiment. This distinction allows a promising candidate to remain in the comparison after a destructive event, provided that optimization is successfully resumed from a safe checkpoint.

For the HPO task studied here, the simulation-design and simulation-runner agents provide the main coordination functions, while planning, job dispatch, health monitoring, recovery, and ranking are executed by deterministic tools. These tools can be invoked either through the agent coordination layer or through a lower-level command-line/API entry for reproducibility checks, debugging, and user-specified batch execution. The latter entry does not replace the agents; rather, it exposes the same deterministic tools when the research-level planning and reporting functions of the agent layer are not needed. Both entry routes use the same shared state and execution rules. Runtime recovery is deliberately restricted to recorded and reproducible actions: the system records the event, chooses a safe checkpoint, writes a modified learning-rate schedule, launches the continuation, and retains the complete recovery history for later ranking and inspection.

The framework release provides the implementation, configuration templates, and execution interfaces needed to reproduce this workflow or apply the same tools to another physical task and search space. The health-monitoring, scoring, and recovery components are included as reusable modules.

\subsection{Optimization Health Diagnosis and Runtime Recovery}
\label{subsec:health_recovery}

The central diagnostic object is the energy trajectory of a candidate experiment. In Fig.~\ref{fig:framework}, runtime monitoring reads the energy log and checkpoint information produced during optimization. A useful trajectory should reach a competitive energy while remaining stable enough to continue. The monitor evaluates rolling-window statistics, the variance and slope near the end of the trajectory, non-finite values, and abnormal fluctuations, including upward deviations from a trusted reference energy. This reference is updated only from a sufficiently stable rolling window: the window standard deviation, normalized by the magnitude of the window mean energy, must not exceed $5\%$. An isolated, implausibly low value is recorded but is not allowed to redefine the reference energy.

We define a destructive optimization event as an instability that requires the current execution segment to stop before optimization can continue. One signature is an upward energy jump that does not recover within the confirmation window. A non-finite energy or loss value, such as NaN or Inf, is classified immediately as a destructive event. Both types can be recovered when a safe checkpoint exists and the allowed number of recovery attempts has not been exhausted. They are distinct from short transient fluctuations that return to the trusted energy range without intervention.

Operationally, the monitor follows a reproducible sequence of checks. When an abnormal fluctuation appears in the energy trajectory, the event is first treated as provisional. The trajectory is then followed over a predefined confirmation interval. If the energy returns to the trusted range and remains stable, the event is classified as a transient fluctuation and the calculation continues without intervention. If the energy does not recover within the confirmation interval, the event is classified as a destructive fluctuation and the recovery procedure is triggered. Non-finite values, such as NaN or Inf, bypass this waiting period and are classified as destructive events immediately. When the fluctuation starts too close to the end of the current evaluation interval to make this distinction reliably, the event is left for later review instead of being assigned a transient or destructive label.

The underlying VMC calculations use the Adam optimizer~\cite{Kingma2015Adam} together with the prescribed learning-rate schedule from the aCNN study~\cite{Jqwang2024aCNN}: the learning rate first increases from zero to the target value and is then reduced at predefined decay milestones. NQS-Agent does not replace this optimizer; instead, it intervenes only when the monitored energy trajectory indicates a destructive event. When such an event is detected at runtime, NQS-Agent stops the current execution segment and searches for the latest safe checkpoint before the event. It then starts a continuation with a modified learning-rate schedule. If the event occurs after the initial increase to the target value, the schedule is modified to multiply the learning rate by $0.5$ shortly before the event. For an event during this initial increase, the maximum learning rate is limited to the pre-event value. The continuation remains part of the same candidate experiment, so later ranking uses the recovered trajectory while retaining the complete instability and recovery history. Recovery is attempted only a limited number of times, with the allowed number of attempts increasing with the target optimization length.

\subsection{Checkpoint-Based Continuation and Anomaly-Aware Ranking}
\label{subsec:ranking}

The search is organized as a staged continuation process rather than as a set of independent fixed-length calculations. Candidates are first run to a prescribed evaluation milestone, defined as a target number of optimization steps at which their energy trajectories and health records are scored. Candidates that pass this comparison are promoted to a later stage with a larger number of optimization steps, whereas less promising or invalid candidates are stopped. This staged allocation of calculation length is related to Hyperband, ASHA, and BOHB, which use short early evaluations to decide which candidates should receive longer follow-up calculations. In the present workflow, however, a promoted NQS candidate continues from a checkpoint that stores both the network parameters and optimizer state. This avoids restarting the candidate as a new calculation and preserves the same optimization trajectory, including the adaptive-optimizer state, checkpoint history, and any recovery events already recorded.

NQS-Agent also supports ASHA and TPE-based search. Those general methods are appropriate when the performance of each candidate can be represented by a single scalar objective, such as the best valid energy reached at a given optimization length. For the aCNN benchmark, candidate selection instead uses a task-specific top-$k$ rule because the decision must consider the energy trajectory, recovery status, and availability of a safe checkpoint in addition to the energy value.

For each physical task, the evaluation milestones are specified in the experiment plan. A checkpoint is a saved state used for continuation, whereas an evaluation milestone is a target optimization length at which candidates are scored and compared. Early milestones compare and identify promising random seeds, learning-rate and architecture combinations; later milestones test the promoted candidates under longer optimization. Within each comparison group, the top-$k$ candidates are selected only after every candidate has completed, stopped with an error, or been cancelled. The evaluation sequence used in the present benchmark around the aCNN reference architecture is summarized in Fig.~\ref{fig:acnn_stage_selection}.

The workflow can also refine the tested learning rates between two evaluations. When the two leading learning rates, $\eta_1$ and $\eta_2$, differ by a sufficiently large multiplicative factor, NQS-Agent can test their geometric mean, $\sqrt{\eta_1\eta_2}$, in an additional confirmation calculation. This choice places the new value midway between $\eta_1$ and $\eta_2$ when the learning rates are compared by their ratio and tests whether a favorable region lies between two coarse grid values.

Ranking uses an anomaly-aware score. The leading term is the lowest median energy ever seen of the rolling window. The remaining terms penalize several signs of unreliable optimization: deterioration near the end of the trajectory, a positive late-time slope, large tail fluctuations, recorded anomaly events, recovery attempts, and incomplete checkpoint status. For a candidate that is allowed to enter the ranking at evaluation milestone $m$, the score is
\begin{align}
S_m =
&E_{\mathrm{trusted}}
+ P_{\mathrm{tail}}
+ P_{\mathrm{var}}
+ P_{\mathrm{slope}} \nonumber\\
&+ P_{\mathrm{anomaly}}
+ P_{\mathrm{recovery}}
+ P_{\mathrm{checkpoint}} ,
\label{eq:anomaly_score}
\end{align}
Here $E_{\mathrm{trusted}}$ is the median energy of the selected trusted window, $P_{\mathrm{tail}}=\max(0,\bar{E}_{\mathrm{tail}}-E_{\mathrm{trusted}})$ measures deterioration of the trajectory end, and $P_{\mathrm{var}}=\sqrt{\mathrm{var}_{\mathrm{tail}}}$ penalizes tail fluctuations. The quantity $s_{\mathrm{tail}}$ is the slope obtained by a linear fit of energy against optimization step within the final rolling window; thus, $P_{\mathrm{slope}}=\max(0,s_{\mathrm{tail}})$ penalizes an overall upward trend near the end of the trajectory. The anomaly penalty $P_{\mathrm{anomaly}}=5\times10^{-4}N_{\mathrm{transient}}+2\times10^{-3}N_{\mathrm{destructive}}$ counts transient and destructive anomaly events with different weights, where $N_{\mathrm{transient}}$ and $N_{\mathrm{destructive}}$ are the corresponding event counts. The recovery penalty $P_{\mathrm{recovery}}=10^{-3}N_{\mathrm{recovery}}$ increases with the number of recovery attempts. The checkpoint penalty is $P_{\mathrm{checkpoint}}=0$ when a safe full-resume checkpoint is available and $P_{\mathrm{checkpoint}}=5\times10^{-4}$ otherwise for a ranking-eligible candidate. A candidate is excluded from ranking rather than merely penalized when the energy log is missing or invalid, a process or resource failure prevents a valid result, a destructive event remains unrecovered, the allowed recovery attempts are exhausted, no complete checkpoint is available for continuation, or the energy is discontinuous immediately after continuation. NaN or Inf alone is not an irreversible failure: it is treated as a destructive event and can enter the recovery procedure when a safe checkpoint and an unused recovery attempt are available.

This ranking rule differs from selecting the lowest final energy at a fixed step. A candidate with a low but unstable energy may be unsuitable for continuation to a larger number of optimization steps. The score therefore requires both a competitive energy and sufficient stability, including the ability to recover safely when an instability occurs.

\section{Results}
\label{sec:results}

\subsection{Benchmark Task, Human Baseline, and Parameter-Count-Matched Search Space}
\label{subsec:benchmark_space}

We benchmark the workflow on the square-lattice spin-$1/2$ Heisenberg $J_1$-$J_2$ model with periodic boundary conditions, a standard benchmark for frustrated magnetism studied by tensor-network, DMRG, VMC, and NQS methods. The main comparison tasks use $10\times10$ lattices at $J_2/J_1=0.50$ and $0.55$, where the reported human-tuned aCNN reference energies are $-0.495627(6)$ and $-0.483490(5)$ per site, respectively~\cite{Jqwang2024aCNN}. The primary observable is the energy per site, with lower values indicating better variational states.

The architecture reported in the aCNN study is denoted A5 and has channel number $c=3$, kernel size $k=5$, residual-block count $b=14$, and 6538 real-valued parameters. To compare different channel, kernel-size, and depth choices at a similar parameter-count scale, we allow a $\pm5\%$ window around 6538 parameters, corresponding to 6211--6864 parameters. About 30 residual-CNN architectures satisfy this condition. The nine representatives in Table~\ref{tab:arch_space} were selected from this eligible set to cover different depth, width, and kernel-size regimes; candidates with kernel size one were not included in the representative subset because they provide little spatial-convolution variation for the present comparison.

\begin{table*}[t!]
  \begin{center}
    \caption{Representative parameter-count-matched residual-CNN architectures. The allowed parameter count is within $\pm5\%$ of the 6538-parameter A5 architecture. Here $c$, $k$, and $b$ denote the channel number, kernel size, and residual-block count, respectively. The nine representatives are selected from about 30 eligible architectures to cover different depth, width, and kernel-size regimes. A5 is the architecture reported in the aCNN study, and A9 is the widest and shallowest representative in this subset.}
    \label{tab:arch_space}
    \begin{tabular}{c c c c c l}
      \hline
      Arch. & $c$ & $k$ & $b$ & Parameters & Qualitative type \\
      \hline
      A1 & 2 & 5 & 32 & 6631 & very narrow, medium kernel, very deep \\
      A2 & 2 & 7 & 16 & 6535 & very narrow, large kernel, deep \\
      A3 & 2 & 9 & 10 & 6847 & very narrow, very large kernel, medium depth \\
      A4 & 3 & 3 & 39 & 6610 & narrow, small kernel, extremely deep \\
      A5 & 3 & 5 & 14 & 6538 & aCNN reference, balanced depth and width \\
      A6 & 3 & 7 & 7 & 6514 & medium width, large kernel, shallow \\
      A7 & 3 & 9 & 4 & 6346 & medium width, very large kernel, very shallow \\
      A8 & 5 & 3 & 14 & 6536 & medium width, small kernel, medium depth \\
      A9 & 11 & 3 & 3 & 6810 & very wide, small kernel, very shallow \\
      \hline
    \end{tabular}
  \end{center}
\end{table*}

\subsection{Multi-Stage HPO and Automatic Recovery from Destructive Events}
\label{subsec:hpo_recovery_results}

\begin{figure*}[t!]
  \includegraphics[width=0.98\textwidth]{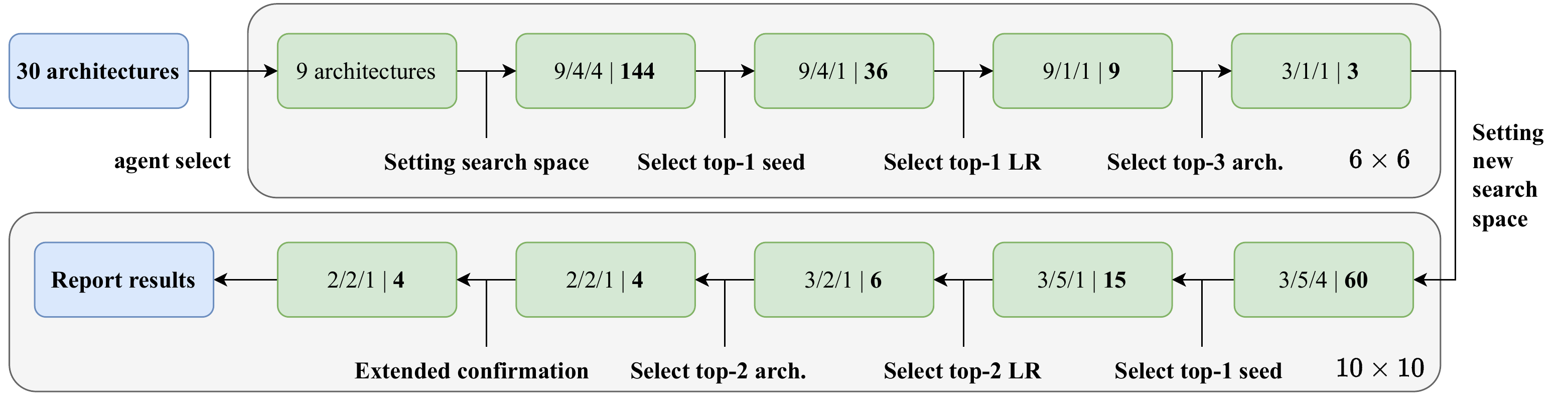}
  \caption{Progressive architecture and learning-rate selection in the present benchmark around the aCNN reference architecture. The calculations for the $6\times6$ lattice compare nine parameter-count-matched residual-CNN architectures using four learning rates and four random seeds and select A5, A9, and A6 for the calculations for the $10\times10$ lattice. At $10\times10$, candidates are evaluated at predefined optimization milestones, compared with the anomaly-aware score, and the candidate set is progressively reduced before longer confirmation calculations. The notation $a/l/s|n$ denotes $a$ architectures, $l$ learning rates, and $s$ random seeds, with $n=a\times l\times s$ executed calculations in the corresponding comparison group.}
  \label{fig:acnn_stage_selection}
\end{figure*}

The HPO process progressively reduces the number of candidates, as summarized in Fig.~\ref{fig:acnn_stage_selection}. The calculations for the $6\times6$ lattice at $J_2/J_1=0.55$ compare nine architectures using four learning rates and four random seeds and select A5, A9, and A6 for subsequent calculations for the $10\times10$ lattice. In the flowchart, $a/l/s|n$ denotes $a$ architectures, $l$ learning rates, and $s$ random seeds, giving $n=a\times l\times s$ executed calculations. The calculations for the $10\times10$ lattice then compare the retained architecture and learning-rate candidates, reassess random-seed outcomes with the anomaly-aware score, select the remaining candidates, and continue them for longer confirmation calculations.

As described in Sec.~\ref{subsec:health_recovery}, NQS-Agent complements the Adam-based optimization protocol used in the aCNN study rather than replacing it. In the examples below, the workflow monitors the observed energy curve, detects destructive events, and, when needed, resumes a promising candidate from a checkpoint with a modified learning-rate schedule.

\begin{figure}[t!]
  \includegraphics[width=8.5cm]{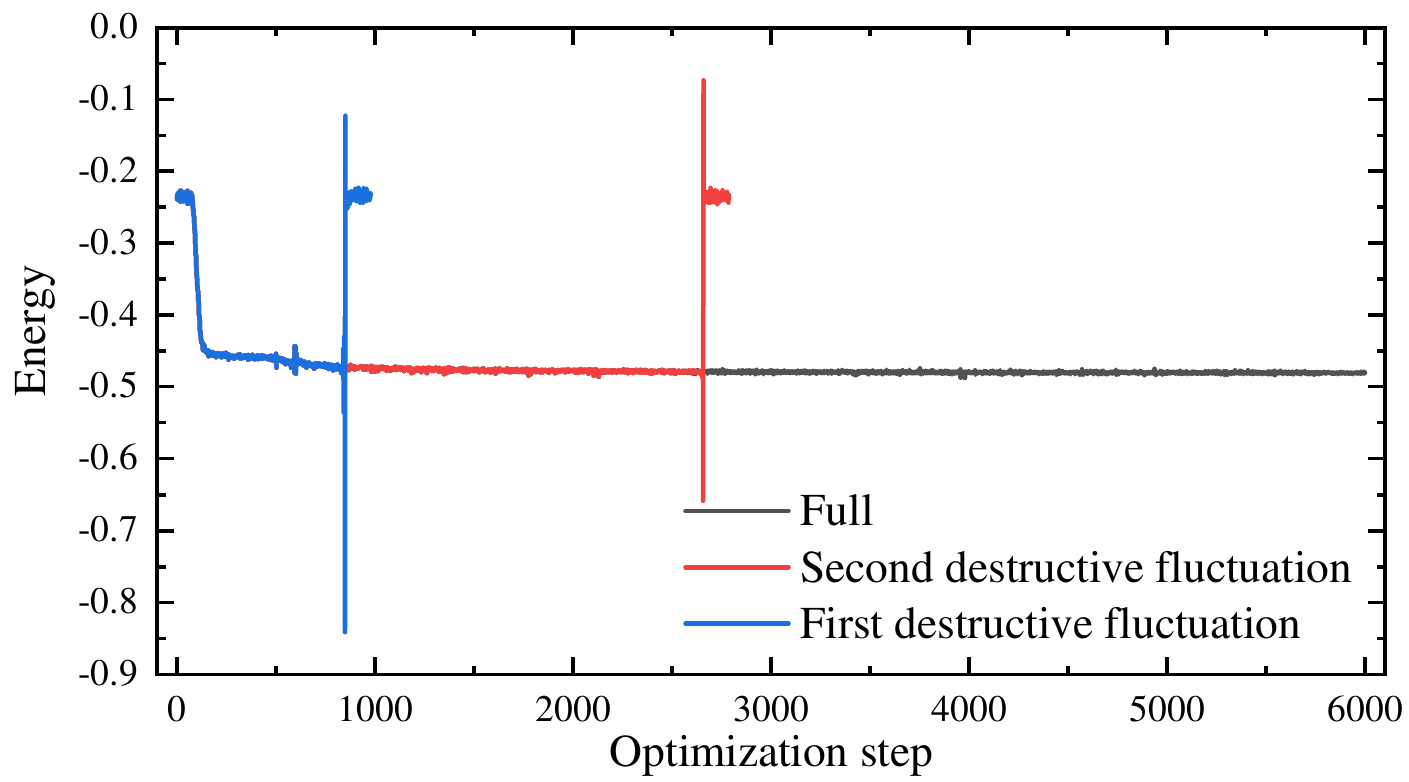}
  \caption{Representative energy trajectory for the $10\times10$, $J_2/J_1=0.55$, A5 calculation with learning rate $0.0013$ and seed 8. This calculation illustrates typical stability and recovery behavior rather than the final selected setting. A non-destructive fluctuation occurs near step 650. Destructive fluctuations are detected in the intervals 840--978 and 2657--2788, after which NQS-Agent stops the unstable execution, resumes from safe checkpoints, and modifies the learning-rate schedule.}
  \label{fig:energy_lineage}
\end{figure}

Figure~\ref{fig:energy_lineage} shows a representative recovered energy trajectory for the $10\times10$, $J_2/J_1=0.55$, A5 calculation with learning rate $0.0013$ and seed 8. A non-destructive fluctuation occurs near step 650 at the end of a rapid energy decrease and does not trigger recovery. In contrast, the upward fluctuations in the intervals 840--978 and 2657--2788 are identified as destructive events. After each confirmed event, the workflow stops the unstable execution segment, selects a safe checkpoint before the event, modifies the learning-rate schedule, and continues the same candidate experiment. This example shows that NQS-Agent acts during optimization rather than only ranking completed calculations: it converts an interrupted candidate into a continued calculation that can still be evaluated.

\begin{figure*}[t!]
  \begin{center}
    \begin{tabular}{cc}
      \includegraphics[width=0.48\textwidth]{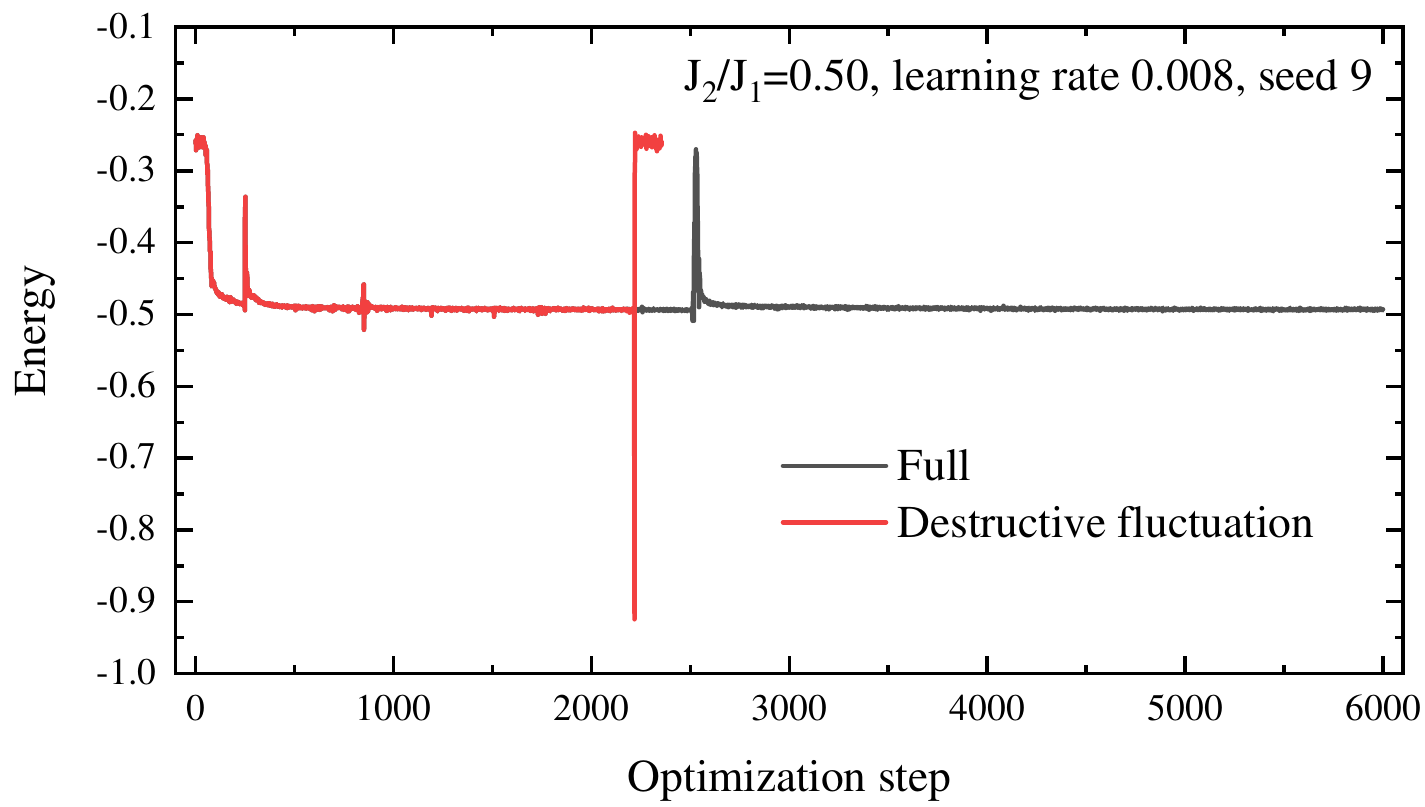} &
      \includegraphics[width=0.48\textwidth]{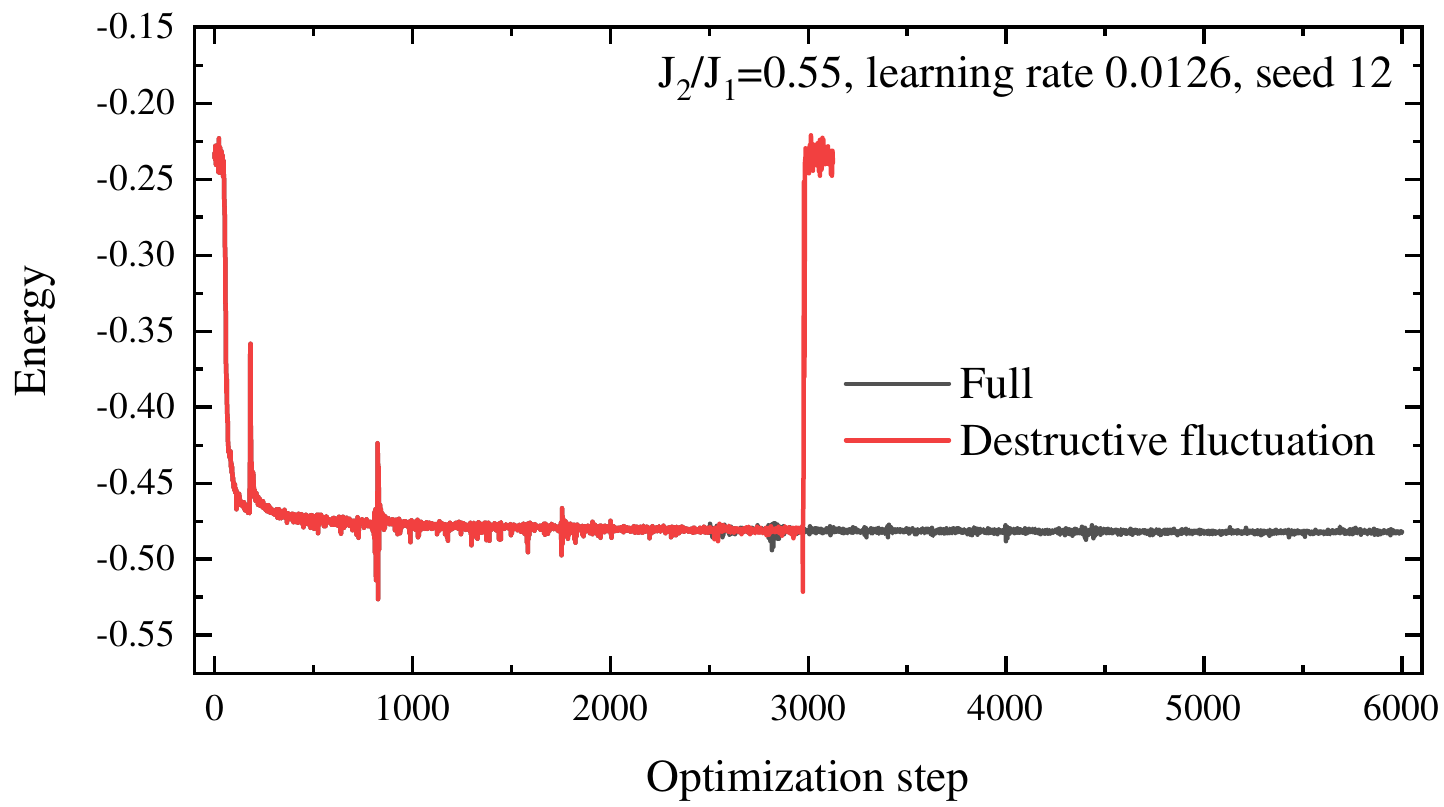}
    \end{tabular}
  \end{center}
  \caption{Energy trajectories for two selected A9 settings in the $10\times10$ parameter-count-matched comparison. Left: $J_2/J_1=0.50$, learning rate $0.008$, seed 9. Right: $J_2/J_1=0.55$, learning rate $0.0126$, seed 12. Both trajectories exhibit destructive fluctuations, yet checkpoint-based recovery with modified learning-rate schedules allows the candidate calculations to continue and subsequently reach energies below the reported human-tuned aCNN baseline.}
  \label{fig:a9_energy_lineages}
\end{figure*}

Figure~\ref{fig:a9_energy_lineages} shows two A9 energy trajectories used in the final parameter-count-matched comparison. Both contain abnormal fluctuations, including destructive fluctuations, but the workflow does not discard the candidates solely because these fluctuations occur. Instead, recovery from checkpoints with modified learning-rate schedules allows the same candidate experiments to continue. With this recovery procedure, candidates that pass through destructive events can subsequently reach energies below the reported human-tuned aCNN baseline.

\subsection{Comparison with Human-Tuned aCNN Results}
\label{subsec:human_comparison}

Table~\ref{tab:hpo_results} summarizes the parameter-count-matched HPO results for the two $10\times10$ physical tasks, including settings that were not continued to the final comparison stage. Among the retained final results, NQS-Agent obtains energies below the reported human-tuned aCNN baseline for both $J_2/J_1=0.50$ and $0.55$. The A5 results directly test automated optimizer tuning and recovery-based continuation for the architecture reported in the aCNN study. The A9 results show that the architecture comparison points to a structurally distinct wide-and-shallow setting with competitive energies. Because the present calculations were not designed as a controlled head-to-head comparison between A9 and A5 across all seeds and optimization settings, A9 is reported as a promising additional architecture rather than as a definitive replacement for A5.

\begin{table*}[t!]
  \begin{center}
    \caption{Parameter-count-matched HPO results on the $10\times10$ Heisenberg $J_1$-$J_2$ tasks. LR denotes learning rate. Energy values are reported only when a calculation was continued to a reported final value. ``Not promoted'' means that the setting was not selected for longer optimization, and ``Failed after promotion'' means that the continued calculation exhausted the allowed recovery attempts without producing a valid final result. The A5 calculation with learning rate $0.001265$ was an additional refinement test at $J_2/J_1=0.55$, obtained as the geometric mean of two selected learning rates; no corresponding test was performed at $J_2/J_1=0.50$, as indicated by ``---''.}
    \label{tab:hpo_results}
    \setlength{\tabcolsep}{15pt}
    \begin{tabular}{l c c}
      \hline
      Arch. and configuration & $J_2/J_1=0.50$ & $J_2/J_1=0.55$ \\
      \hline
      A5, LR=$0.005$, seed 8 & $-0.495892(6)$ & Not promoted \\
      A5, LR=$0.0013$, seed 8 & Not promoted & $-0.482319(10)$ \\
      A5, LR=$0.002$, seed 9 & $-0.495959(5)$ & $-0.483502(7)$ \\
      A5, LR=$0.001265$, seed 8 & --- & $-0.483208(6)$ \\
      \hline
      A9, LR=$0.008$, seed 9 & $-0.495941(6)$ & Not promoted \\
      A9, LR=$0.020$, seed 10 & Failed after promotion & $-0.483528(8)$ \\
      A9, LR=$0.0126$, seed 12 & Not promoted & $-0.483512(5)$ \\
      \hline
      aCNN human-tuned baseline & $-0.495627(6)$ & $-0.483490(5)$ \\
      \hline
    \end{tabular}
  \end{center}
\end{table*}

The A5 calculation with learning rate $0.001265$ illustrates the optional learning-rate refinement in which NQS-Agent tests an intermediate value between two selected learning rates. It tests the geometric mean of two selected learning rates for the $J_2/J_1=0.55$ A5 comparison. Although it does not give the lowest A5 energy in Table~\ref{tab:hpo_results}, it demonstrates that NQS-Agent can add a targeted follow-up calculation rather than remaining limited to the initial learning-rate grid.

The A9 results have a different implication. A9 is not the architecture reported in the aCNN study; it is a structurally distinct residual CNN with many more channels, a smaller kernel, and far fewer residual blocks. Its competitive energies, including values below the aCNN baseline at both tested coupling ratios, show that the parameter-count-matched residual-CNN space contains promising candidate alternatives to the human-selected architecture. This outcome is noteworthy because it does not support a simple assumption that a deeper residual CNN is always the most reliable route to improved variational accuracy. Within the present search space, A9 should be regarded as a promising architecture identified by the HPO workflow, while A5 remains the appropriate reference architecture for the direct comparison between automated and human tuning.

\subsection{Architecture-Dependent Optimization Behavior and Workflow Efficiency}
\label{subsec:architecture_behavior}

Architecture hyperparameters strongly affect both optimization behavior and practical computational cost. We therefore compare the static complexity estimates, measured execution speeds, and later HPO outcomes of the nine parameter-count-matched architectures. A1--A4 are not selected for the subsequent calculations for the $10\times10$ lattice, and A6 is selected but does not dominate the later HPO results. The balanced A5 architecture remains the primary setting for testing the HPO protocol, whereas the wide-and-shallow A9 architecture emerges as a competitive alternative. Comparable parameter counts do not imply equal computational cost, because the architectures differ substantially in residual-block count, stored activations, and practical execution efficiency.

\begin{table}[t!]
  \begin{center}
    \caption{Static complexity estimates for the nine residual-CNN architectures in the calculations for the $6\times6$ lattice. The reported Forward FLOPs and Stored activations are normalized to A5. Forward floating-point operations (FLOPs) estimate the arithmetic work in one network forward pass. Stored activations estimate the amount of intermediate layer output stored during training so that gradients can be computed.}
    \label{tab:architecture_cost}
    \setlength{\tabcolsep}{2.0mm}{
      \begin{tabular}{c c c}
        \hline
        Arch. & Forward FLOPs & Stored activations \\
        \hline
        A1 & 1.008 & 1.467 \\
        A2 & 1.003 & 0.756 \\
        A3 & 1.055 & 0.489 \\
        A4 & 0.988 & 2.667 \\
        A5 & 1.000 & 1.000 \\
        A6 & 1.003 & 0.533 \\
        A7 & 0.980 & 0.333 \\
        A8 & 0.991 & 1.667 \\
        A9 & 1.044 & 0.978 \\
        \hline
      \end{tabular}%
    }
  \end{center}
\end{table}

These quantities are architecture-level estimates: Forward FLOPs count arithmetic work in the network evaluation, whereas stored activations approximate the intermediate layer outputs that must be retained for gradient computation. They are calculated from the architecture rather than measured with a performance-analysis tool, and stored activations should not be interpreted as direct measurements of peak GPU memory.

\begin{figure}[t!]
  \includegraphics[width=8.5cm]{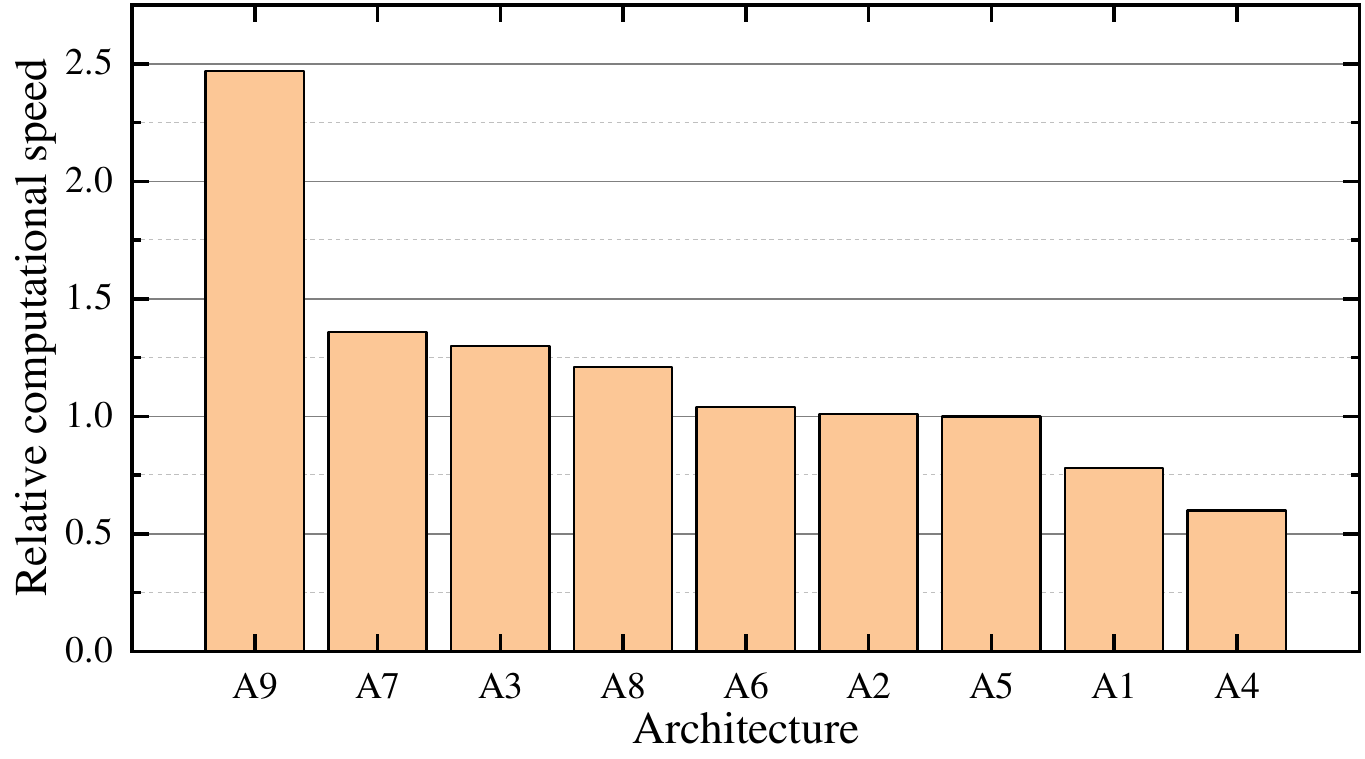}
  \caption{Relative computational speed of the nine architectures in the calculations for the $6\times6$ lattice, normalized to A5. Relative speed is calculated from the average computation time required for one optimization step under the same $6\times6$ calculation protocol, with values above one indicating faster execution than A5. All timing measurements were performed on an NVIDIA RTX 4090 GPU.}
  \label{fig:relative_speed_6x6}
\end{figure}

The forward FLOPs vary only from about $0.98$ to $1.06$ times the A5 value, whereas the stored activations span approximately $0.33$--$2.67$ times the A5 value. Figure~\ref{fig:relative_speed_6x6} shows that the measured speed ordering is not determined by Forward FLOPs alone. The two deepest candidates, A4 and A1, have the largest stored-activation values and the lowest measured relative speeds. Shallower architectures such as A7, A3, and A6 have lower stored-activation values, although their measured speeds do not follow the FLOPs ordering. Most notably, A9 has forward FLOPs and stored activations close to A5 but is approximately $2.47$ times as fast as A5 in the measured $6\times6$ timing test, whereas A4 is only about $0.60$ times as fast as A5. These differences show that parameter count and static FLOPs alone do not predict practical execution speed; residual-block count, architecture-dependent parallelism, the overhead of launching many layers, and details of the aCNN implementation can also contribute.

Computational efficiency and optimization outcome are related but distinct criteria. A3 and A7 execute faster than A5 but are not selected for the later calculations for the $10\times10$ lattice, showing that faster execution does not necessarily imply better optimization or variational performance. A5 remains a strong balanced architecture, whereas A9 is notable because it combines the highest measured relative speed with competitive variational results in the later calculations for the $10\times10$ lattice. A8 provides a similar example that faster execution does not necessarily coincide with selection by the energy- and stability-based HPO procedure. These observations support reporting architecture-dependent computational cost alongside energy and optimization stability in HPO studies.

These trends should not be interpreted as universal statements about depth, width, or speed in NQS. They apply to the residual-CNN family, the parameter-count window around 6538 parameters, and the calculations considered here. Nevertheless, they show why architecture hyperparameters belong in HPO: a human-selected architecture can be strong, while architectures in the same parameter-count window can have competitive accuracy and substantially different computational costs.

Workflow efficiency also has an organizational component beyond architecture-dependent computational cost: NQS-Agent improves the management and traceability of large HPO studies. For each candidate, it records the simulation settings, comparison group, saved checkpoints, detected health events, recovery decisions, and ranking status. These records make it possible to determine why a candidate was continued for additional optimization, stopped, recovered, or excluded from ranking. Such traceability turns a large collection of HPO calculations into a reproducible scientific workflow.

\section{Discussion}
\label{sec:discussion}

The main contribution of this work is a shift from selecting NQS candidates only by energy to health-aware HPO. General HPO methods can evaluate many candidate settings, but NQS optimization requires additional information about the energy trajectory. A candidate that briefly reaches a low energy can still be unsuitable if it has no safe checkpoint, ends with an unrecovered destructive fluctuation, or becomes discontinuous immediately after continuation from a checkpoint. Conversely, a promising candidate that undergoes an abnormal fluctuation should not necessarily be discarded when the calculation can be safely recovered.

The aCNN benchmark separates two effects. The A5 results test whether automated optimization of learning-rate settings can improve a human-selected reference architecture. The A9 results test whether varying structural hyperparameters within the same parameter-count window can reveal a qualitatively different candidate. These results require different claim strengths: A5 directly evaluates automated tuning against human tuning for the same architecture, whereas A9 shows that the same HPO protocol can identify promising structural candidates under controlled comparison conditions.

Several limitations remain. First, the architectures are matched by parameter count, not by total computational cost. Although the results on the $6\times6$ lattice reveal substantial architecture-dependent differences in relative speed, the measured ratios may depend on the software environment and details of the aCNN implementation. Second, the current recovery rules are deterministic and deliberately conservative. More flexible agent-based decisions may help with ambiguous fluctuations near the end of a trajectory, but such decisions must remain subject to explicit safety rules. Third, the present results concern residual CNN architectures based on the aCNN study; the generality of the protocol should therefore be assessed with additional NQS ansatz classes and physical problems.

More broadly, the results indicate that optimization health should be assessed together with energy when deciding whether an NQS candidate is reliable and suitable for continuation. By integrating this assessment with recorded recovery and selection decisions, NQS-Agent makes automated NQS studies easier to audit and reproduce, providing a practical basis for broader automation in this research area.

\section{Summary}
\label{sec:conclusion}

We introduced NQS-Agent as an implemented open-source software framework for health-aware HPO of neural-network quantum states. The central motivation is that an NQS candidate should not be judged only by the energy reached at a selected evaluation point: the stability of the energy trajectory, destructive fluctuations, checkpoint availability, and recovery history also determine whether the result is reliable and suitable for continuation. NQS-Agent supports large-scale automated searches over NQS optimization and architecture parameter spaces by combining trajectory monitoring, anomaly-aware ranking, checkpoint-based recovery, and learning-rate schedule adjustment within a recorded and reproducible workflow. Applied to parameter-count-matched residual CNNs for the square-lattice Heisenberg $J_1$-$J_2$ model, the workflow improves over the reported human-tuned result for the aCNN reference architecture and identifies A9 as a structurally distinct competitive candidate. These results show that health-aware HPO provides a practical protocol for NQS tuning and architecture comparison. More broadly, NQS-Agent provides an implemented framework and basic toolset for reproducible automated research in NQS.

\begin{acknowledgments}
This work was supported by the National Natural Science Foundation of China (Grant No. 12434009) and the National Key R\&D Program of China (Grants No. 2024YFA1408602 and No. 2024YFA1408601). Z.-Y.L. was also supported by the Innovation Program for Quantum Science and Technology (Grant No. 2021ZD0302402). Computational resources were provided by the Physical Laboratory of High Performance Computing in Renmin University of China.
\end{acknowledgments}

\section*{Code availability}

The NQS-Agent source code \footnote{Repository URL: https://github.com/QTMEC-RUC/NQS-Agents} will be made publicly available on GitHub upon publication. The release will include the implementation shown in Fig.~\ref{fig:framework}, configuration templates, and example CLI/API entry points for running and reproducing the workflow.
\bibliography{reference}

\end{document}